\documentclass{elsart3p}
\usepackage{graphicx,microtype}
\usepackage{txfonts,color, natbib}
\journal{New Astronomy}

\begin{document}
\newcommand{\mps}{m\,s$^{-1}$}
\newcommand{\arcsec}{$^{\prime\prime}$}
\newcommand\apj{{ApJ{ }}}%
\newcommand\apjl{{ApJ{ }}}%
\newcommand\ao{{Appl.~Opt.{ }}}%
\newcommand\aap{{A\&A{ }}}%
\newcommand\mnras{{MNRAS{ }}}%
\newcommand\solphys{{Sol.~Phys.{ }}}%
\newcommand\nat{{Nature{ }}}%

\begin{frontmatter}

  \title{Large-scale horizontal flows in the solar photosphere}
  
  \subtitle{IV. On the vertical structure of large-scale horizontal flows}

  \author[asu,au]{M.~\v{S}vanda\corauthref{cor}},
  \corauth[cor]{Corresponding author.}
  \ead{michal@astronomie.cz}
  \author[asu]{M.~Klva\v{n}a},
  \ead{mklvana@asu.cas.cz}
  \author[asu]{M.~Sobotka},
  \ead{msobotka@asu.cas.cz}
  \author[hepl]{A.~G.~Kosovichev}, and 
  \ead{sasha@quake.stanford.edu}
  \author[nasa]{T.~L.~Duvall,~Jr.}
  \ead{thomas.l.duvall@nasa.gov}
  
  \address[asu]{Astronomical Institute, Academy of Sciences of the Czech Republic (v.~v.~i.), Fri\v{c}ova 298, CZ-251~65 Ond\v{r}ejov, Czech Republic}
  \address[au]{Astronomical Institute, Faculty of Mathematics and Physics, Charles University, V Hole\v{s}ovi\v{c}k\'{a}ch 2,
             CZ-180~00 Prague~8, Czech Republic}
  \address[hepl]{W. W. Hansen Experimental Physics Laboratory, Stanford University, Stanford, CA~94305, USA}
  \address[nasa]{Solar Physics Laboratory, NASA Goddard Space Flight Center, Greenbelt, MD~20771, USA}

  \begin{abstract}
  {In the recent papers, we introduced a method utilised to measure the flow field. The method is based on the tracking of supergranular structures. We did not precisely know, whether its results represent the flow field in the photosphere or in some sub-photospheric layers. In this paper, in combination with helioseismic data, we are able to estimate the depths in the solar convection envelope, where the detected large-scale flow field is well represented by the surface measurements. We got a clear answer to question what kind of structures we track in full-disc Dopplergrams. It seems that in the quiet Sun regions the supergranular structures are tracked, while in the regions with the magnetic field the structures of the magnetic field are dominant. This observation seems obvious, because the nature of Doppler structures is different in the magnetic regions and in the quiet Sun. We show that the large-scale flow detected by our method represents the motion of plasma in layers down to $\sim$10~Mm. The supergranules may therefore be treated as the objects carried by the underlying large-scale velocity field.}
  \end{abstract}
  
  
  \end{frontmatter}
   
\section{Introduction}

The supergranulation is a convection-like pattern that remains a puzzle since its discovery by \cite{1954MNRAS.114...17H}. There are many papers published examining the parametric properties of supergranular cells. Typical values of the sizes are around 20--30~Mm, lifetime of 24~hours \citep[see e.g.][ and references herein]{2004ApJ...616.1242D}. There is no explanation yet to give a clear statement about the origin of supergranules. The classical mechanism invoked to explain the origin of the supergranulation is the latent heat of the recombination of He$^{2+}$ into He$^+$ at roughly 10~Mm below the photosphere \citep{1985ApJ...288..795G}. In a fluid at rest, such a heat release may trigger an instability, which can turn into motions at scales comparable to the supergranular scale. The issue in the Sun is that the plasma in the convection zone is highly turbulent and the mentioned thermal instability can be suppressed by turbulent motions. There is also a lot of works doubting the existence of the supergranulation as the convection mode. The surface properties of the supergranules may be explained as non-linear interaction between granules triggered by exploding granules \citep[e.g.][]{2000AA...357.1063R}. The wave-like properties of the supergranulation were also discussed \citep[e.g.][]{2003Natur.421...43G}.

The time-distance local helioseismology \citep{1993Natur.362..430D} is a tool that allows to examine the structure of the plasma flows beneath the solar surface. The $p$-modes of the solar oscillations are excited in the upper convection zone and travel through the convective envelope. Plasma motions and variations of the sound-speed change the speed of the wave-packet propagation. From the difference between the theoretical travel-time from one point in the solar photosphere to another one and the theoretically calculated one, one can map the disturbances causing the travel-time deviations in the subphotospherical layers. Initial results of the time-distance helioseismology from SOHO/MDI \citep{1997SoPh..170...63D} showed that a pattern of the surface flows can persist 2--3~Mm below the surface, improved results gave 8~Mm \citep[][]{1998ESASP.418..581D}. Further time-distance analyses suggested the depth of the supergranular flow to be $\sim15$~Mm \citep{2003ESASP.517..417Z}. In that study, the authors calculated the correlation of the horizontal divergence at different depths with the horizontal divergence just below the surface. They found a positive correlation in the depths of 0--5~Mm. Deeper down it changes the sign. It may mean that the return flow in the supergranules starts to dominate the intercellular flow structure.

The supergranules are well visible on the entire solar disc in Dopplergrams. This is caused by the prevailing horizontality of the internal velocity field within the supergranular cells. If we assume that supergranules are objects carried by the flow field on the larger scale, the visibility of supergranules in the Dopplergrams gives an opportunity to use them to map this underlying large-scale flow field. It was first done by \cite{1990ApJ...351..309S}. Recently, we developed a method based on the local correlation tracking algorithm \citep[LCT; ][]{1986ApOpt..25..392N} with a similar utilisation. This method \citep[for details see][]{2006AA...458..301S} uses the pattern of the supergranulation in the SOHO/MDI Dopplergrams \citep{1995SoPh..162..129S} to measure the motions on larger scales. The issue of the method is that we measure displacements of the supergranular structures in the series of the processed Dopplergrams and interpret them as the large-scale velocity field in the solar photosphere. However, we cannot establish the range of depths in the solar convection zone, where the large-scale flows are well represented by the surface measurements. With the use of the data provided by the time-distance helioseismology we can determine this unknown parameter. We can also verify the assumption of our analysis: whether the supergranules are subjected to the transport by the velocity field on the larger scales.

\begin{figure}[!b]
\centering
\resizebox{0.40\textwidth}{!}{\rotatebox{0}{\includegraphics{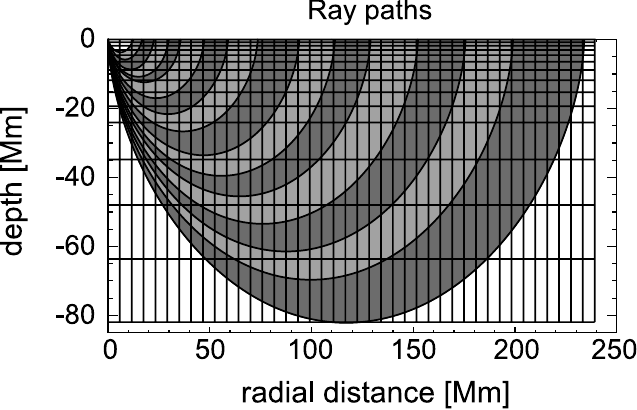}}} 
\caption{A sample of acoustic ray paths used for the time-distance helioseismology, shown in a vertical plane. The shadowed regions illustrate ranges of averaging. The vertical and horizontal lines show a grid used for inversion of acoustic travel time data. After \cite{2006SSRv..124....1K}, who used the same dataset as in this study.}
\label{fig:fig5}
\end{figure}

There is a list of works suggesting that the supergranules do not represent the actual differential rotation or meridional circulation. \cite{2000SoPh..193..333B} mentioned that rotation rate of the supergranules determined from MDI Dopplergrams is 5\,\% larger than the corresponding spectroscopic rate. Some explanations of this phenomenon appeared in the following papers. \cite{2003Natur.421...43G} suggested the modulation of the convective pattern by travelling waves. Already \cite{1995ApJ...443..423W} noted that the effective viscous dissipation of Rossby-waves ($r$-modes), which happen in the Sun at 0.932~$R_\odot$ ($\sim$50~Mm below the photosphere), could modulate the convection in the near-subphotospherical layer. \cite{2006ApJ...644..598H}, however, used synthetic data to explain the apparent super-rotation of the supergranulation as the projection effects on the line-of-sight Doppler velocity signal. \cite{2007AA...466..691M} re-analysed Hathaway's idea and found that the projection effects can explain the most of the supergranular super-rotation, but not all of it. \cite{2007AA...466..691M} concluded that the supergranules are unreliable in the analysis of the motions on supergranular scales. However, in \cite{2007SoPh..241...27S} the results showed that the large-scale apparent motion of supergranules measured by the method introduced in \cite{2006AA...458..301S}, and used also in this study, is in almost perfect agreement with the near-surface mass flow measured by the time-distance helioseismology. It is perhaps a $k$--$\omega$ filter used for the suppression of the noise, which makes this method more useful than the previous ones based on similar principles. 

The large-scale flows are a combination of many types of motions on various spatial scales (such as the differential rotation, meridional circulation, possibly the giant cell convection, and others), which unfortunately cannot be reliably separated in components. These components may vary independently with time and depth. This implies that, in principle, when dealing with the large-scale flows, we cannot obtain unambiguous results that do not allow any alternative interpretation. Nonetheless, the realistic numerical simulation should reveal the combined large-scale plasma flow, properties of which should be comparable with the measurements. Such numerical simulation, which is not present at the time, shall allow to distinguish various components of the detected large-scale flow and to study them separately. The results in this paper therefore provide encouraging large-scale flow properties, which are to be reproduced by models.

\section{Data and Method}
For the comparison between the data obtained by our method and the results of the time-distance helioseismology we used high-cadence full-resolution full-disc Dopplergrams recorded in March and April 2001 by SOHO/MDI. In this period, 46 flow maps of the vicinity of the active region NOAA~9393/9433 averaged over 8~hours were computed by the time-distance helioseismology using the ray approximation \citep[see][]{2004ApJ...603..776Z}. From these maps we used five examples in five different days (March 3rd, March 28th, 29th, 30th, and April 25th). In these days, MDI data series did not contain many gaps, so the data were suitable for our tracking method, and the field-of-view was far from the solar limb, so we can exclude any possible limb effect.

The one-day series of full-disc Dopplergrams were processed following the procedure introduced by  \cite{2006AA...458..301S} with a few modifications. First, the Postel projection was used instead of the Sanson-Flamsteed one, because helioseismic data are computed in this geometrical projection (it conserves the main circles, therefore is suitable for the measurement of $p$-modes travel-times). In \cite{2006AA...458..301S} we performed the computations in two main steps. The first step was used for the determination of the differential rotation profile, which was then removed during the remapping of the data to decrease the maximum displacement caused by motions of supergranular structures. In the second step, the flow field with respect to the removed mean rotation profile was calculated using the algorithm with a higher precision. Then the differential rotation profile calculated in the first step was added to the flow field calculated in the second step. In this study the computations are done in one step, i.e., no differential rotation is removed before the application of the finer LCT algorithm. We assume that in a small field-of-view, the change in the differential rotation profile is not important. 

In summary, the data processing routine consists of the $p$-modes removal (using the weighted average over 30 minutes), coordinate alignment (all frames record the same region on the Sun), coordinate transformation (using the Postel projection) and a $k$--$\omega$ filtering with the cut-off of 1500\,\mps{} for the noise removal. Then, the LCT algorithm for the determination of supergranular structures displacements is used to calculate the 24-hours averaged large-scale velocity field with the 60\arcsec{} correlation window.

As we discussed in \cite{2006AA...458..301S}, our method provides the measurements with a random-error of 15~\mps{}. It is believed that the time-distance helioseismology errors are within 10~\% in the near-surface layers and generally increase with depth, under some 20~Mm the details of the results must be considered unreliable \citep{Zhao2006priv}. Unfortunately, it is impossible to determine the actual errors using the ray approximation. The study made by \cite{2004ApJ...616.1261B} showed that the estimated errors in the supergranular or moat flow achieved by both ray and Born approximations are of the order of 10~\%. The development of the helioseismic method is fast and we can expect more reliable results with exactly determined accuracies using a perturbation theory and fully consistent OLA inversions \citep{2008SoPh..251..381J} in a few years.

All the datasets were aligned with the centre of the field-of-view on Carrington coordinates $l=148.5^\circ$ and $b=19^\circ$. The field-of-view has the total size of 512$\times$512 pixels with the resolution of 0.92\arcsec{}\,px$^{-1}$. The datacubes from the local helioseismology contain 15 irregularly spaced depths from 0.77~Mm to more than 80~Mm. The vertical resolution is based on the ray approximation. The numbers rather than isolated levels represent the consecutive intervals in the radial direction, in which the results are averaged. Fig.~\ref{fig:fig5} shows examples of the acoustic ray paths, which clearly illustrates the vertical resolution.

\begin{figure*}[!t]
\resizebox{0.49\textwidth}{!}{\rotatebox{0}{\includegraphics{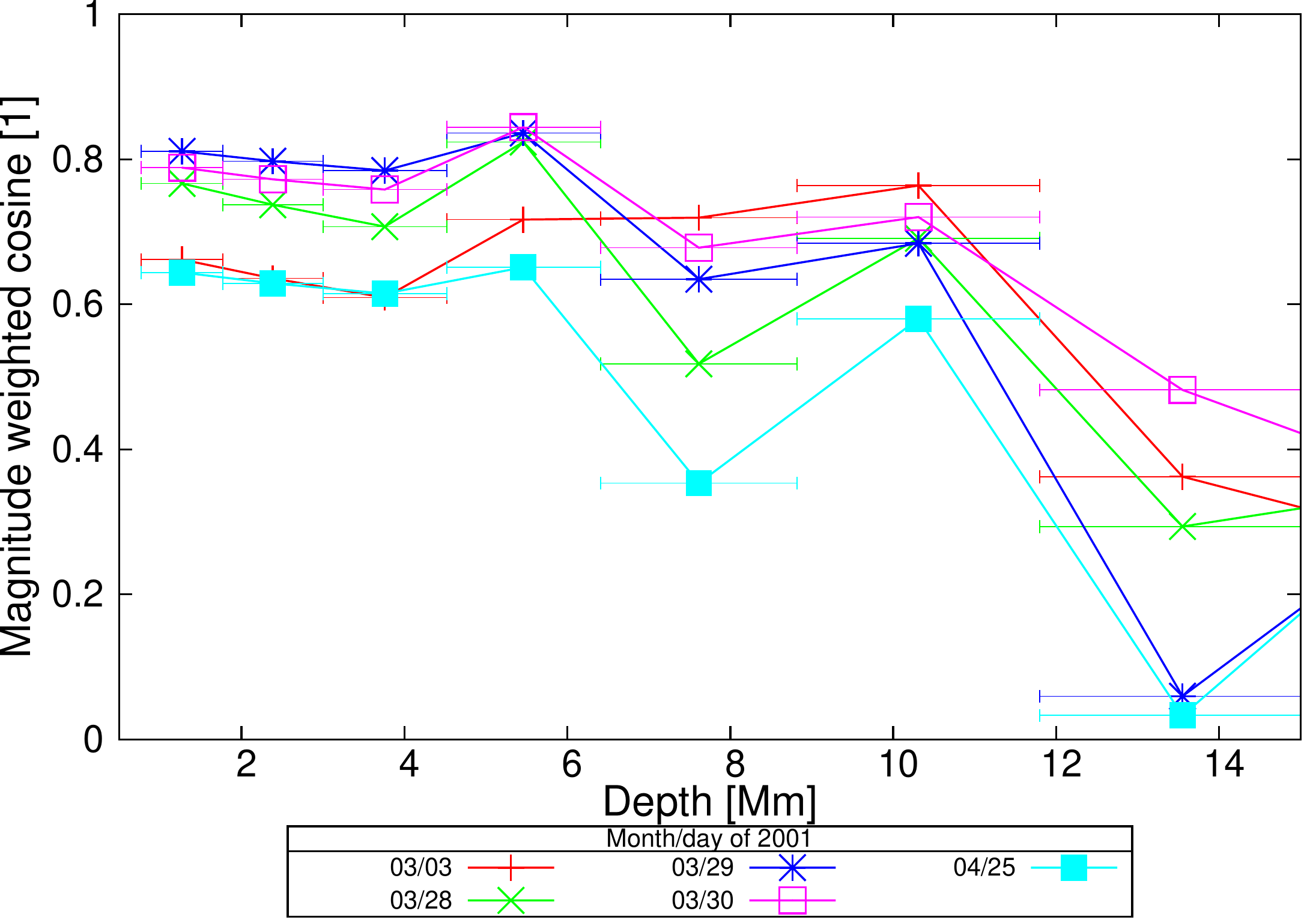}}} \resizebox{0.49\textwidth}{!}{\rotatebox{0}{\includegraphics{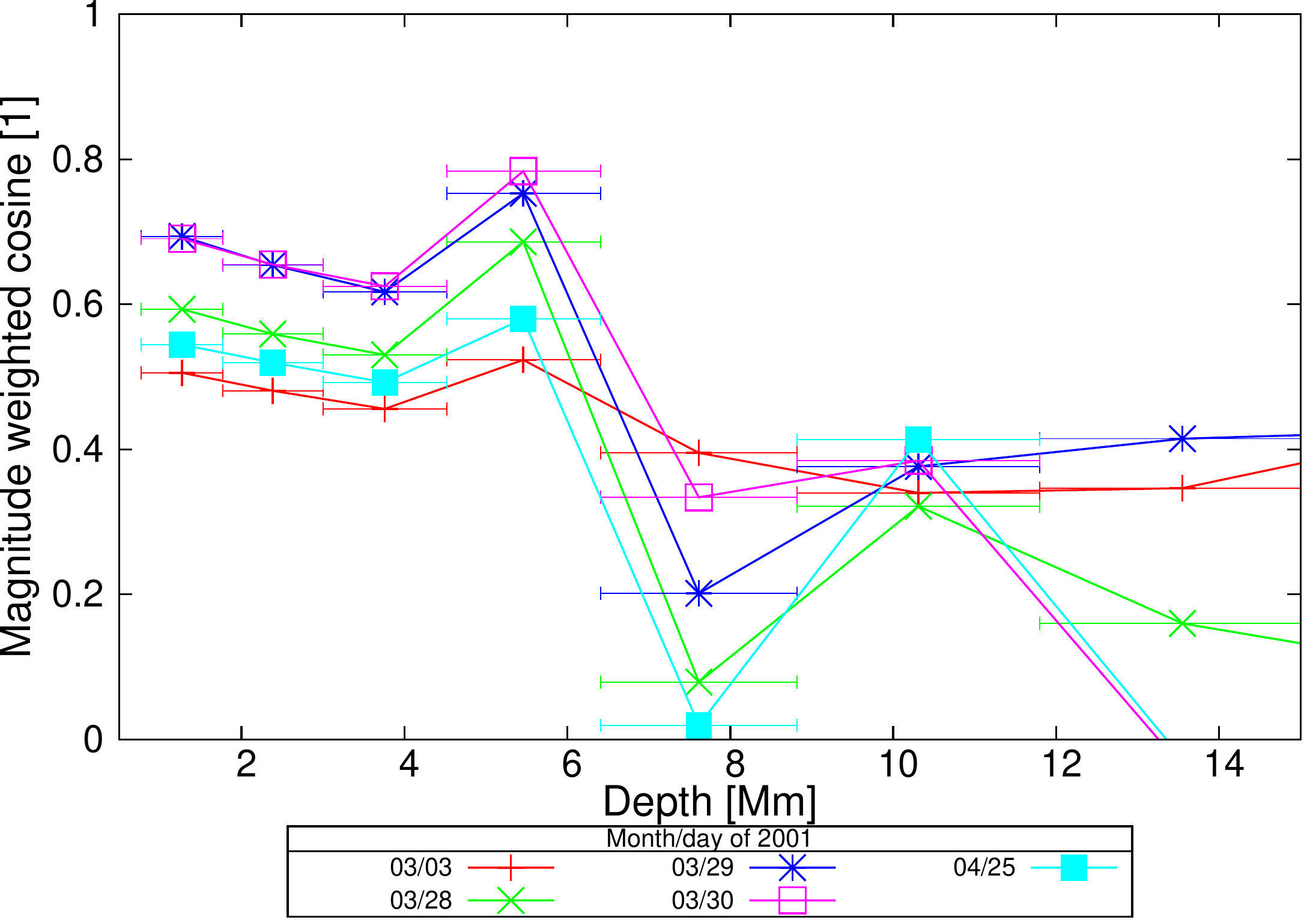}}}
\caption{The dependence of the similarity of the velocity fields provided by the time-distance helioseismology and the flows calculated by the LCT method on the depth in the solar convection zone. \emph{Left:} in quiet Sun regions, \emph{right} in the regions occupied by the magnetic field.}
\label{fig:fig1}
\end{figure*}

The results of our tracking method have an effective resolution of 60\arcsec{}, but the results of the time-distance helioseismology have 8\arcsec. We binned the helioseismic flow maps to match the resolution of 60\arcsec. This also means that the structure of the internal flow in the supergranulation is filtered out and does not disturb the study of the large-scale flow field. This is the most important deviation from the study of \cite{2003ESASP.517..417Z}. 

The velocity field obtained by our tracking method is in principle twodimensional -- horizontal. The $z$-component of the time-distance flows suffers from effects of cross-talks, therefore we use only horizontal ($x$ and $y$) components. For each depth in the helioseismic datacube we calculated its similarity to the flow map obtained by the tracking of supergranular structures. As the measure of the similarity we used the magnitude weighted cosine of the direction difference \citep[as in][]{2007SoPh..241...27S}, which is robust to the presence of the noise. This quantity is given by the formula:
 
\begin{equation}
\rho_{\rm W} = \frac{\sum |\vec{a}| \frac{\vec{a} \cdot \vec{b}}{|\vec{a}||\vec{b}|}}{\sum |\vec{a}|},
\end{equation}
where $\vec{a}$ and $\vec{b}$ are vector fields, $\vec{a} \cdot \vec{b}$ is a scalar multiplication and $|\vec{a}|$ is a magnitude. The closer this quantity is to 1, the better the alignment between two vector fields. In our case, $\vec{a}=\vec{v}_{\rm LCT}$ and $\vec{b}=\vec{v}_{\rm time\mbox{-}distance}({\rm depth})$. The magnitude weighted cosine is also stable to any possible issue involving the magnitude determination error in the time-distance inversion.

We investigate the similarity of both methods results separately for the regions occupied by the magnetic field and the regions of the quiet Sun. Two regimes (magnetic and non-magnetic) are separated in the field-of-view applying the masks calculated from the surface MDI magnetograms. We smoothed the magnetograms by 20\arcsec{} and estimated the threshold of 5~Gauss in the smoothed magnetograms as the level distinguishing the magnetic regions from the non-magnetic ones. 

\begin{figure*}[!t]
\resizebox{0.9\textwidth}{!}{\includegraphics{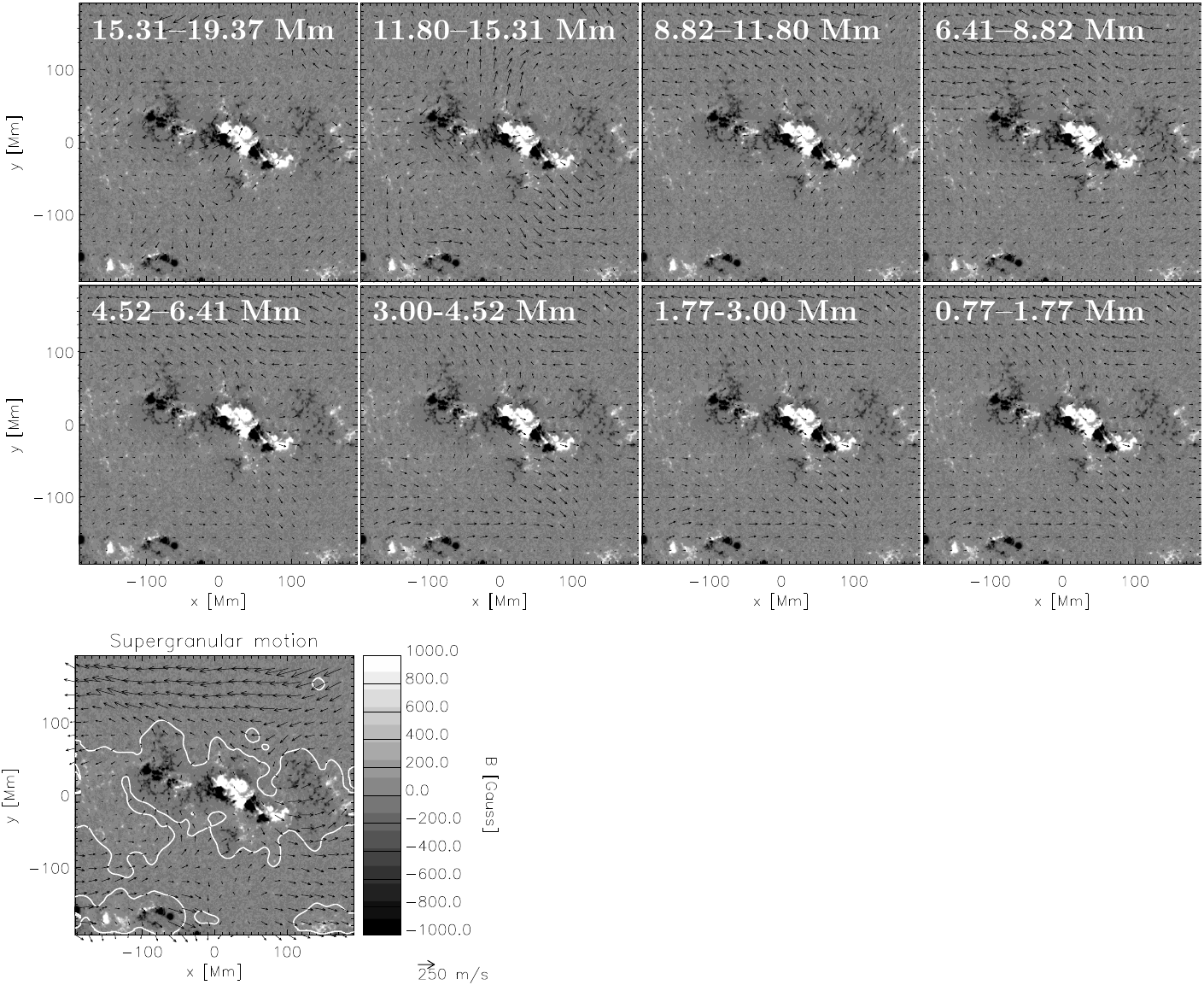}}
\caption{The examples of the flow maps in the vicinity of active region NOAA~9393 on March 29th 2001. We see that the topology of flows measured by the local helioseismology is similar with the surface one measured by LCT down to depth of roughly 10~Mm, below that the topology changes. The lengths of arrows in the time-distance flow maps are not in the same scale, their magnitude increases with depth. In the ``Supergranular motion'' subframe the contours of the mask distinguishing the magnetic and non-magnetic regions are overplotted.}
\label{fig:fig2}
\end{figure*}

\begin{figure*}[!t]
\resizebox{0.9\textwidth}{!}{\includegraphics{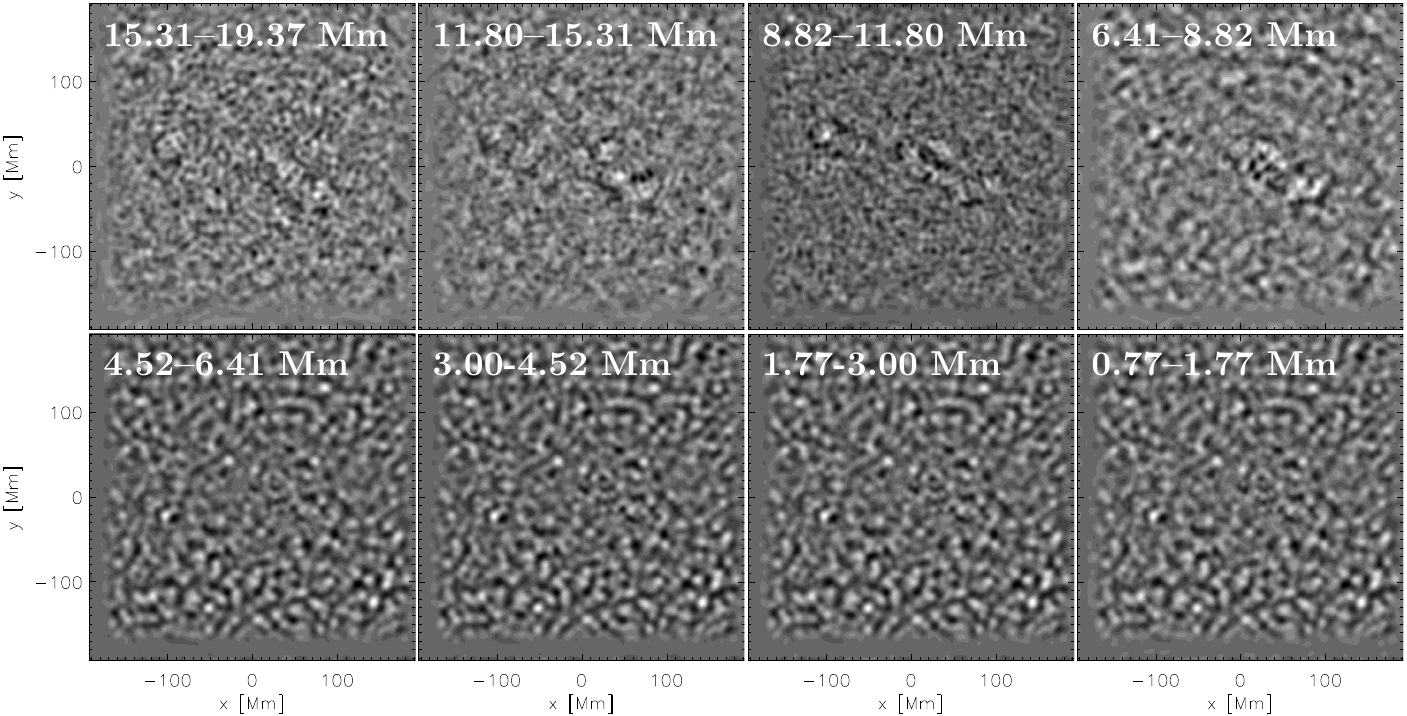}}
\caption{Horizontal divergence maps of the internal supergranular flow at various depths in the solar convection zone. Dark shades of grey denote inflows, while light shades represent outflows, shades of grey are scaled automatically for an easy comparison of the structure between each other.}
\label{fig:fig3}
\end{figure*}
\section{Results}

\subsection{Large-scale horizontal flows}

The results of the comparison are displayed in Fig.~\ref{fig:fig1}. We basically see a large similarity of the velocity fields measured by both methods at higher levels of the solar convection zone. The maximum is reached in all but one studied cases at the level of 4.5--6.4~Mm ($\sim 0.006~R_\odot$). For the quiet Sun regions, the absolute value of the magnitude weighted cosine is not significantly different in the layer 4.5--6.4~Mm and in the neighbouring layer 3.0--4.5~Mm, it is within the estimated 10\% accuracy.

Our analysis shows that the large-scale flows do not vary much with depth down to 10~Mm in the quiet Sun regions. This fact means that the large-scale horizontal velocity field measured on the surface represents well the large-scale horizontal dynamics in the layers of 0--10~Mm in the depth. See Fig.~\ref{fig:fig2} for examples of the flow maps at different depths. We cannot expect much better match, because both methods measure different properties. While LCT measures the motion of structures, the time-distance helioseismology maps the plasma flow. Both results may be very different if any wave-like phenomenon is involved. 
   
In magnetised regions, the dominant structures in the Dopplergrams are the structures connected rather to the magnetic field than to the supergranulation. This idea is supported by our results, where we see that in the magnetised regions the detected flows are coherent within the depths down to $\sim$5~Mm with the peak in the depths of 4.5--6.4~Mm. This is the depth, where downward convergent flows are still detected is sunspots \citep[e.g.][]{2001ApJ...557..384Z} and where sunspots stop exist as the regions of the suppressed heat transport. The existence of the Evershed flow in sunspots, which is easily seen in Dopplergrams, supports our interpretation of the velocity field in magnetised regions, too. 

There is an alternative interpretation of the obtained results. Assuming that the time-distance helioseismology is less reliable deeper down in the convection zone and that the large-scale flows do not vary much in the upper convection zone, our results could put the level, where the reliability of the time-distance measurements is reasonable. In this interpretation, the time-distance flow measurements are reasonable in upper 10~Mm in the quiet Sun regions and upper 5~Mm in regions occupied by the magnetic field. There are recent papers \citep[e.g.][]{2008SoPh..251..381J} using another time-distance implementation, which suggest that the current time-distance results are reasonable only in the upper-most 3~Mm.

\subsection{Vertical structure of the supergranulation}

The obtained results can be interpreted as a fulfilment of the basic assumption of our tracking method: The supergranular structures may be indeed treated as the objects carried by the large-scale velocity field. The supergranules move in the depths down to 5~Mm coherently in the horizontal direction with a slight loss of the coherence between 5 and 10~Mm in the depth. To confirm this interpretation, the structure of the supergranules itself must be analysed.

Therefore we modified our analysis and instead of the smearing the time-distance flow field to remove the signal of the supergranular internal mass flow, we subtracted the large-scale velocity field, so that only the internal velocities in supergranules remained in the flow map. From the maps containing only the separated supergranulation signal we calculated the horizontal divergence (see Fig.~\ref{fig:fig3}) as the representation of down/up-flows. It is more stable than the measured $z$-component of the velocity, which suffers of plenty of issues \citep{Zhao2006priv}. 

\begin{figure}[!t]
\resizebox{0.49\textwidth}{!}{\rotatebox{0}{\includegraphics{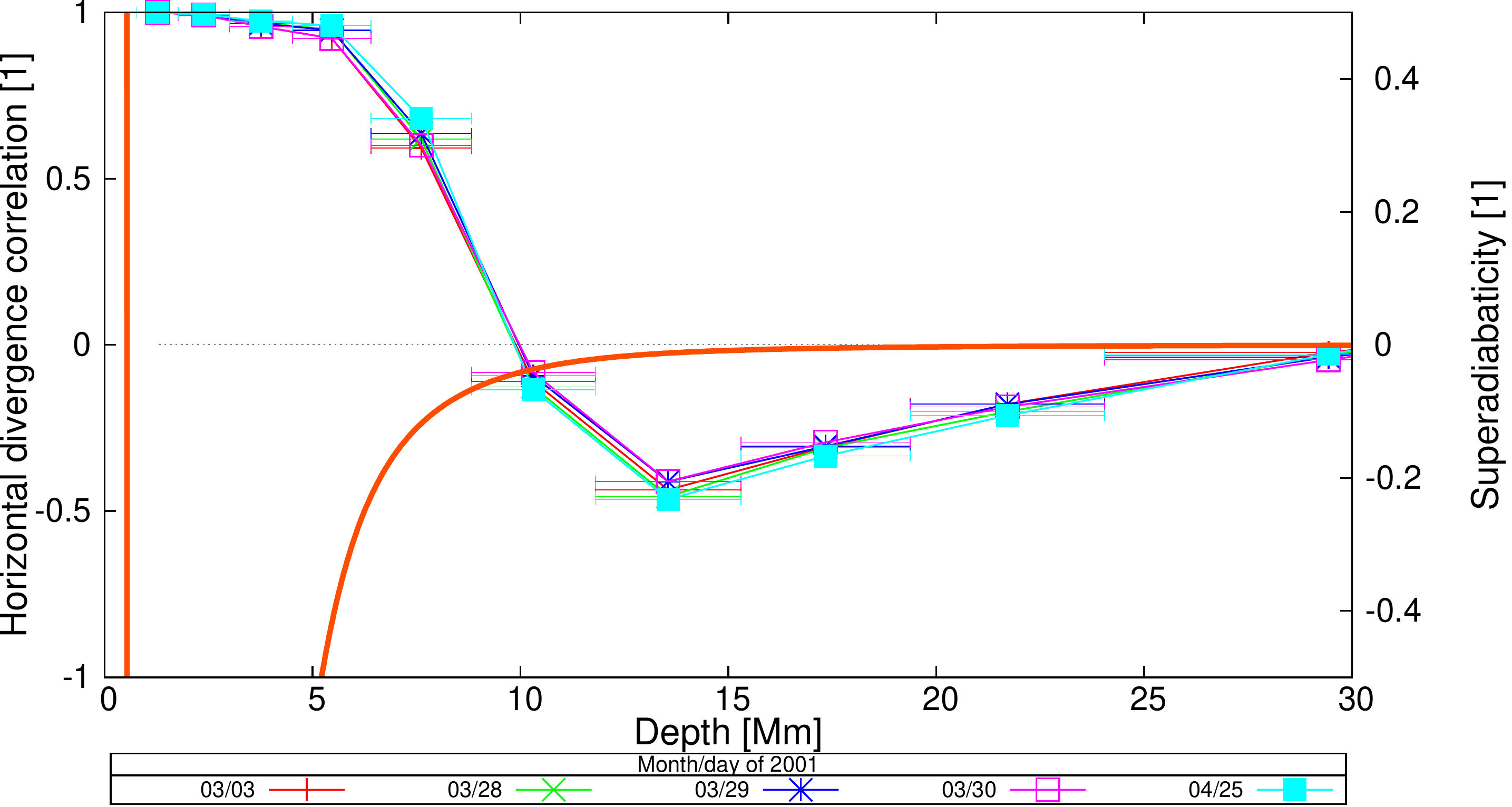}}} \caption{Correlation of divergence maps from Fig.~\ref{fig:fig3} at various depths with the surface layer in regions of quiet Sun. With the thick line, the superadiabaticity in the layers calculated from the model~S \citep{1996Sci...272.1286C} is overplotted.}
\label{fig:fig4}
\end{figure}

From the sequence in Fig.~\ref{fig:fig3} we see that the divergence signal is very similar in the quiet Sun regions in the depths 0.77--6.4~Mm. It is confirmed when plotting the correlation of the near-surface divergence map with maps at various depths (see Fig.~\ref{fig:fig4}). Down to $\sim$7~Mm the correlation is positive, implying the coherence in the structure of the supergranules. Deeper down the correlation turns negative, which might suggest the evidence of the return flow in supergranules. The typical spatial scale of divergence structures goes down by the factor of 2--3 in layers deeper than $\sim$8~Mm. A very similar analysis was done by \cite{2003ESASP.517..417Z} on another dataset with not much different results.

From our data, we cannot compare the structure of the supergranulation between the surface measurement (the Dopplergram) and the top-most helioseismic layer (0.77--1.77~Mm). This topic was, however, assessed by \cite{2000JApA...21..339G}. Authors of that paper demonstrated the very high correlation between the measured surface Dopplergram and the reconstructed line-of-sight velocity image from the time-distance flows in the depths of 0--2~Mm. Based on that and this paper we may conclude that the supergranules are highly coherent in the depths of 0--6.4~Mm.

In Fig.~\ref{fig:fig4} the trend of the superadiabaticity within the sub-surface layers of the convection zone coming from the reference solar model~S \citep{1996Sci...272.1286C} is overplotted. The superadiabaticity $A^*$ is defined as
\begin{equation}
A^*(r)=\frac{1}{\gamma(r)} \frac{{\rm d} \ln p(r)}{{\rm d} \ln r}-\frac{{\rm d} \ln \rho(r)}{{\rm d} \ln r}\ ,
\end{equation}
where $\gamma$, $p$, and $\rho$ are the state parametres (adiabatic exponent, pressure, and density) of the plasma at distance $r$ from the centre of the Sun. The layers are convectively unstable, where $A^*(r)$ is negative \citep[see discussion in][]{1997ApJ...474..790D}. It is to be noticed that $A^*$ turns more negative at the depths of ~10--12~Mm, where it is assumed that the supergranulation should be formed, and remain very negative up to the solar photosphere. It is the same layer, where the large-scale flow is coherent. This observation may be interpreted that the supergranules, which form at the depth some 10~Mm below the surface, are carried by the large-scale velocity field, which operates here (and perhaps deeper). Therefore the large-scale flows remain nearly constant within upper 10~Mm of the convection zone and are detected using our tracking method in surface measurements. 

\section{Conclusions}

Let's put two separate results obtained in the previous Section together. Using the comparison of the large-scale horizontal flows obtained by our method based on the tracking of the supergranular structures and by the local helioseismology we state that the results of the tracking method represent the horizontal dynamical behaviour in the depths down to $\sim$10~Mm below the solar surface. We believe that this result may be important for those, who develop numerical models of the large-scale convection. 

The supergranulation is coherent within layers, which show correlated large-scale horizontal flows. The interpretation of supergranules is closer to the real convective cells carried by the underlying velocity field. The depth of the loss of coherence in large-scale horizontal flows coincides with the layer, where the supergranules as a strong pattern should be formed. 

The results on the vertical structure of the supergranulation we show in this paper, however, cannot be presented as the general results for the supergranulation all over the Sun. We have to keep in mind that these results were obtained only using one suitable dataset in the quiet Sun regions near the very large active region. Much more datasets are needed to describe the behaviour of the supergranulation at different stages of the magnetic activity. Perhaps the helioseismic data expected from the Solar Dynamics Observatory (SDO) will provide some new clues in this topic.

The ray approximation used in the computation of the helioseismological data is often in doubts, suggesting that better (e.g. Born approximation) approach is necessary. It is known \citep[e.g.][]{2001ApJ...561L.229B} that the ray approximation underestimates the small-scale perturbations. The real perturbations inside the Sun are likely to be stronger than those inferred using the ray theory. Inversions, however, involve complicated averages of observed travel times and are regularised. As the result, the details of the inaccuracies of the ray approximation on ray-based inversions are not yet clear. Some comparative studies \citep[e.g.][]{2001ApJ...553L.193J} comparing the results of the ray and Born approximation approaches showed that for the large-scale flow they reasonably match. It was argued that the ray approximation could give credible results with fewer computations. The structure of the flow on supergranular scales was tested in \cite{2001ApJ...557..384Z} with satisfying results.

At this point we cannot exclude that our results instead of bringing the new information in the behaviour of the supergranules rather put limits on the reliability of the time-distance helioseismology. If this interpretation applies, then the time-distance measurements of the large-scale flows are reliable within upper 10~Mm in the quiet Sun regions and within upper 5~Mm in the magnetised areas. Only the improvement of the time-distance helioseismology methods can resolve this ambiguity.

\begin{ack}
M.~\v{S}, M.~K., and M.~S. were supported by the Grant Agency of Academy of Sciences of the Czech Republic under grant IAA30030808, M.~\v{S} additionally by ESA-PECS under grant 98030. The Astronomical Institute of ASCR is working on the Research project AV0Z10030501 (Academy of Sciences of CR), the Astronomical Institute of Charles University on the Research program MSM0021620860 (Ministry of Education of CR). SOHO is a project of international cooperation between ESA and NASA.

\end{ack}

\newcommand{\SortNoop}[1]{}

\end{document}